\documentclass[aps,prb,twocolumn,superscriptaddress,showpacs,amsmath,amssymb,footinbib,longbibliography]{revtex4-2}
\usepackage[english]{babel}
\usepackage{graphicx}
\usepackage{dcolumn}
\usepackage{bm}
\usepackage{subfigure}
\usepackage{color}
\usepackage{textcomp}
\usepackage{float}
\usepackage{siunitx}
\usepackage{soul}

\usepackage[colorlinks,bookmarks=false,citecolor=blue,linkcolor=red,urlcolor=blue]{hyperref}

\newcommand{\xref}[1]{\protect\ref{#1}}
\newcommand{\figref}[1]{Fig.~\protect\ref{#1}}

\renewcommand{\eqref}[1]{Eq.~(\protect\ref{#1})}
\newcommand{\secref}[1]{Sec.~\protect\ref{#1}}

\newcommand{\qop}[1]{\hat{#1}} 
\newcommand{\qvecspin}[1]{\hat{\vec{S}}_{#1}} 

\newcommand{\HM}[1]{\textcolor{blue}{#1}}

\begin{document}

\title{Magnetocaloric properties of centered molecular quantum spin systems}
\author{Hamza Meel}
\email{hmeel@physik.uni-bielefeld.de}
\author{J\"urgen Schnack}
\email{jschnack@uni-bielefeld.de}
\affiliation{Fakult\"at f\"ur Physik, Universit\"at Bielefeld, Postfach 100131, D-33501 Bielefeld, Germany}

\begin{abstract}
We investigate the magnetocaloric properties of a class of centered magnetic molecules that
are very similar in their magnetic properties. In particular, we study the magnetocaloric response
of these molecules in the space of two exchange parameters and as function of the spin quantum 
numbers. Major figures of merit such as adiabatic temperature change and isothermal entropy change 
as well as theoretically achievable low temperatures show that overall 
ferromagnetic interactions are preferential as long as this does not lead to dipolar ordering.
\end{abstract}

\maketitle

\section{Introduction}
\label{sec:introduction}

Magnetic molecules are promising materials for magnetocaloric applications.
Current research efforts focus on the major figures of merit such as adiabatic 
temperature change and isothermal entropy change as well as theoretically 
achievable low temperatures \cite{EvB:DT10}. These properties largely depend 
on two ingredients: (1) the intra-molecular properties as determined
by the Zeeman diagram of the molecule and (2) the properties of the 
three-dimensional lattice of molecules in the solid sample that are
governed by the longe-range dipolar interaction.

\begin{figure}[ht!]
\centering
\includegraphics*[width=0.40\columnwidth]{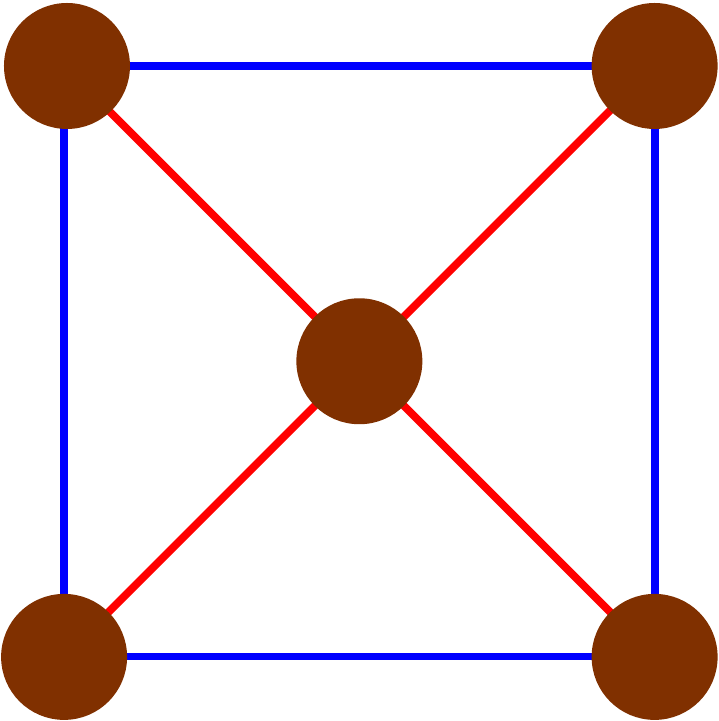}

\includegraphics*[width=0.55\columnwidth]{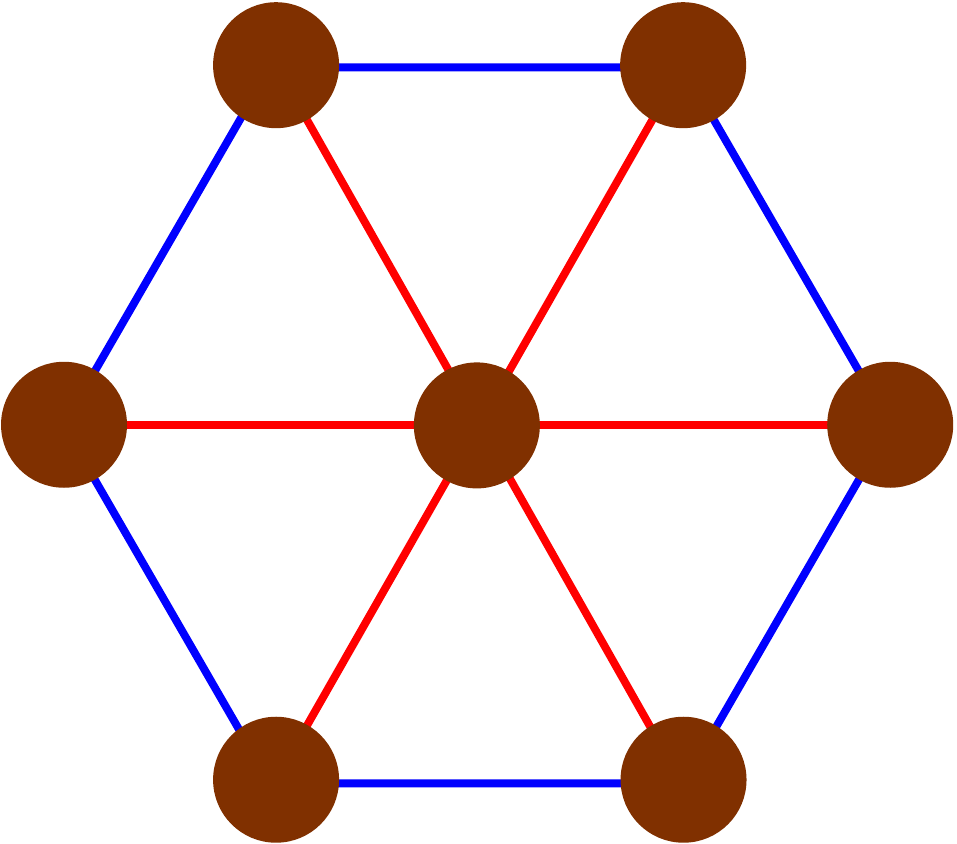}

\includegraphics*[width=0.55\columnwidth]{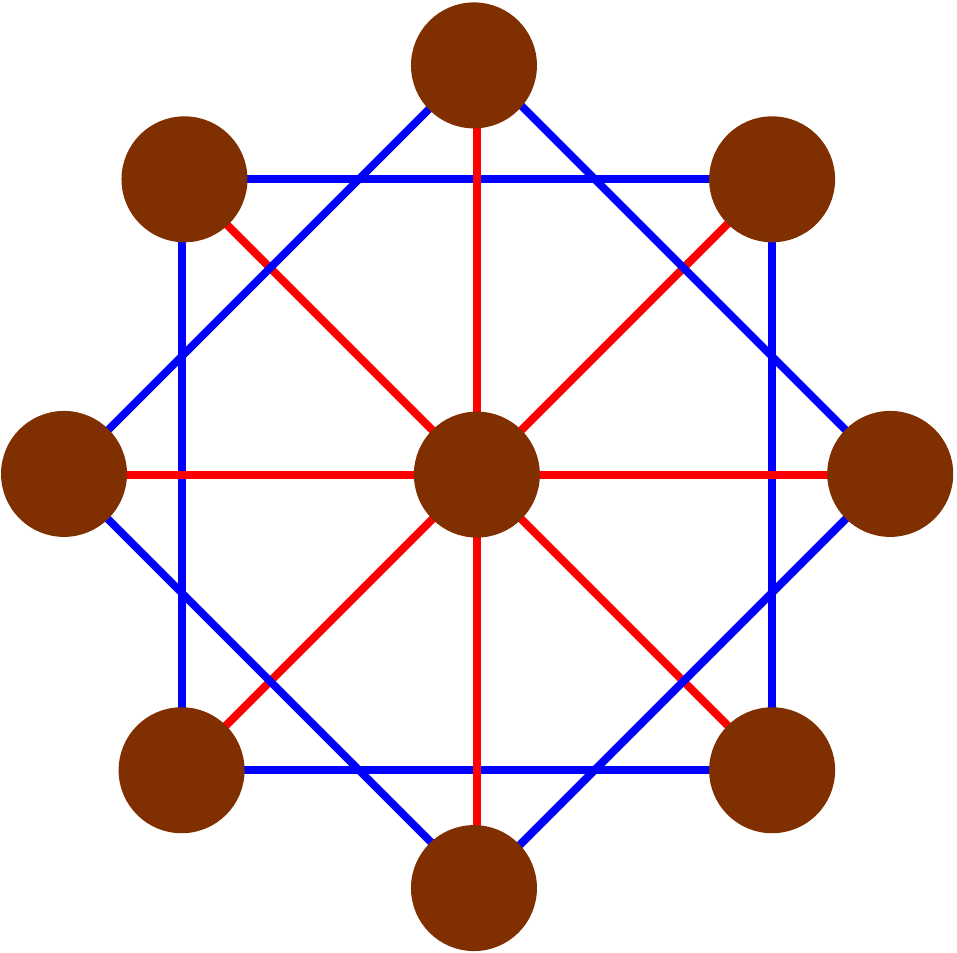}
\caption{Investigated molecular structures: centered square, centered hexagon, centered double pyramid. 
Spins interact via two Heisenberg exchange interactions: $J_1$ (blue) on the squares and the hexagon, 
$J_2$ (red) center spin with all other spins.}
\label{tetracaloric-f-1}
\end{figure}

In this paper, we address the magnetocaloric properties of 
three realistic molecular 
structures as depicted in \figref{tetracaloric-f-1} (top to bottom): 
a centered square, a centered hexagon, and a centered double pyramid.
We investigate these molecules in the Heisenberg model as function of
two exchange interactions that are shown by different colors in 
\figref{tetracaloric-f-1}. It was noted, that molecules with specific antiferromagnetic 
exchange interactions do exhibit unusual quantum effects of decreasing
and increasing adiabatic temperature upon lowering the external field
driven by a strongly varying low-lying density of states 
due to geometric frustration \cite{SCM:NC14,PMM:JACS25}.

Similar to earlier investigations \cite{GCS:APL13,ELP:APL14,GCS:APL14} we
scan the parameter space of the two assumed exchange interactions $J_1$ and $J_2$
in order to characterize the molecular magnetocaloric properties
with a particular interest towards sub-Kelvin cooling with initial
temperatures in the kelvin range.
In summary, 
the results qualitatively show that
(1) paramagnetic materials, i.e., uncoupled spins do perform well
as is known after many decades of their application in real adiabatic 
demagnetization refrigeration (ADR),
(2) ferromagnetic interactions that yield large molecular ground-state spins 
provide a large residual entropy at zero temperature and thus
may also lead to good magnetocaloric properties.
However, since the unavoidable dipolar interaction leads to magnetic ordering
at low temperatures, which prevents further cooling, a final statement
on the MCE properties always has to include investigations of the 
three-dimensional realization. Numerical investigations can incorporate 
intra-molecular dipolar interactions \cite{WeS:JLTP26}, however due to the
long range nature of dipolar interactions, calculations of the quantum system
are virtually impossible.

Geometric frustration of the molecular exchange interactions, 
on the other hand, may thus be a favourable alternative 
since it leads to a large low-lying molecular density of states 
\cite{HoW:PB06,Sch:DT10,LWH:pssb13,SVR:PRB22,RSH:ZNA24},
however, of smaller molecular ground state spins. 

The paper is organized as follows. Section~\xref{sec:model} introduces the model and the quantities of interest,
Sec.~\xref{sec:results} presents and discusses numerical results, 
and Sec.~\xref{sec:discussion} concludes with a summary.

\section{Model}
\label{sec:model}

In the present study, we model the magnetic molecules subject to 
an external magnetic field 
(supposed to be directed along the $z$-axis) using a spin Hamiltonian 
with Heisenberg exchange interactions and a Zeeman term
\begin{align}\label{eq:mce-hamop}
	\qop{H}=&\sum_{i<j} J_{ij} \qvecspin{i}\cdot\qvecspin{j}+ g\mu_B  B \sum_i \qop{S}_i^z
	\ ,
\end{align}
where $\qvecspin{i}$ denotes the spin vector operator at site $i$.
Hats are used to denote operators, and $g = 2$. With this convention, 
a negative interaction strength $J_{ij}$ results in a ferromagnetic interaction, 
and $J_{ij} > 0$ in an anti-ferromagnetic interaction.

The molecular structures studied are presented graphically in \figref{tetracaloric-f-1}. 
They are the centered square, the centered hexagon, and the double pyramid (or hourglass)
shown from top to bottom in \figref{tetracaloric-f-1}. 
Note that the Hamiltonian only depends on the topology of the interactions, 
not on the positions of the sites in space, so the first two structures can also 
be viewed as pyramids with a square or hexagonal base, respectively. 
The structures all follow the same pattern: Without the central spin site the frame 
would not be frustrated if all interactions would be anti-ferromagnetic. 
With a central site that interacts with all other sites, the total spin system 
is geometrically frustrated for all interactions being antiferromagnetic.

We denote the interaction strength among frame-sites $J_1$ 
and the interaction strength between the frame-sites and the central site $J_2$. 
The Hamiltonian can then be expressed as
\begin{eqnarray}
    \qop{H}(J_1, J_2, B) 
    &=& J_1\sum_{i<j \in \text{frame}} \qvecspin{i} \cdot \qvecspin{j} 
    \\
    &+& J_2 \qvecspin{\text{center}} \cdot \sum_{i \in \text{frame}} \qvecspin{i} 
    + g\mu_B  B \sum_i \qop{S}_i^z
    \nonumber
    \ .
\end{eqnarray}

The quantities of interest are a) the change of entropy when turning off the external magnetic field 
(from $B=\SI{7}{\tesla}$ to $B=0$) at constant temperature $T$ called the isothermal entropy change $\Delta S(T)$ 
and b) the temperature reached after adiabatic demagnetization of a sample that was originally at temperature $T_\text{hot}$, $T_\text{cold}(T_\text{hot})$. Both can be computed if the entropy $S(T, B)$ is known. 
The isothermal entropy change is simply $\Delta S(T) := S(T, B=0) - S(T; B=\SI{7}{\tesla})$. $T_\text{cold}(T)$ 
is the solution for $T_*$ in the equation $S(T_*, B=0) = S(T; B=\SI{7}{\tesla})$. 
Note however that this equation actually does not have a solution if $S(T=0,B=0) > S(T, B=\SI{7}{\tesla})$. 
In that case, we take $T_\text{cold}$ to be \SI{0}{\kelvin}.

The entropy of the system is computed using the canonical ensemble, which first requires 
finding the eigen-energies of the Hamiltonian. For the centered square and double pyramid 
they are computed analytically (see Appendix \xref{sec:appendix-analytical-solutions}), 
while for the centered hexagon, they are computed numerically for the case $B=0$ using 
exact diagonalization and leveraging both $U(1)$ and $C_6$ symmetry to speed up computations. 
Since the Hamiltonian is $SU(2)$ symmetric when $B=0$, the eigenvalues of $\qop{H}(J_1,J_2,B=0)$ 
and their corresponding $M$ quantum number can be used to compute the eigenvalues 
of $\qop{H}(J_1, J_2, B)$ for arbitrary $B$. Moreover $\qop{H}(\lambda{}J_1, \lambda{}J_2,B=0) = \lambda\qop{H}(J_1,J_2,B=0)$. 
Hence for the centered hexagon we decided to diagonalize the Hamiltonian for 50 points on the $J_1$-$J_2$ 
unit circle and use the two previous results to compute the energy spectrum for the case 
$\sqrt{J_1^2 + J_2^2} \neq \SI{1}{\kelvin}$.

\section{Results}
\label{sec:results}

We will discuss how the isothermal entropy change $\Delta S$ and 
the temperature reached after adiabatic demagnetization $T_\text{cold}$ 
depend on the temperature $T$ and the interaction strengths $J_1$ and $J_2$. 
By temperature, we mean the temperature of the process for $\Delta S$ and the temperature before demagnetization for $T_\text{cold}$. $\Delta S$ is given per number of sites to make a fair comparison 
between the different molecular structures. 
$J_1$, $J_2$, $T$, and the spin of each site makes for a daunting parameter space, 
and as such some choices had to be made to keep this article readable.

We will first discuss what happens when all sites have spin 3/2 (\secref{sec:equal-spins}) 
and then discuss what happens when the central spin is lowered to 1/2 (\secref{sec:lower-central-spin}). 
We also focus on $T = \SI{10}{\kelvin}$ and $T = \SI{2}{\kelvin}$ since they are good limits 
of the behaviour when changing the temperature. An animation showing how $\Delta S(T)$ evolves when $T$ is changed is available in the supplementary material.

The behavior for other choices of spin quantum numbers is qualitatively the same.

\subsection{Equal spins}
\label{sec:equal-spins}

\subsubsection{Isothermal Entropy Change}
\label{sec:isothermal-entropy-change}

\begin{figure}[H]
    \centering
    \includegraphics[width=0.80\columnwidth]{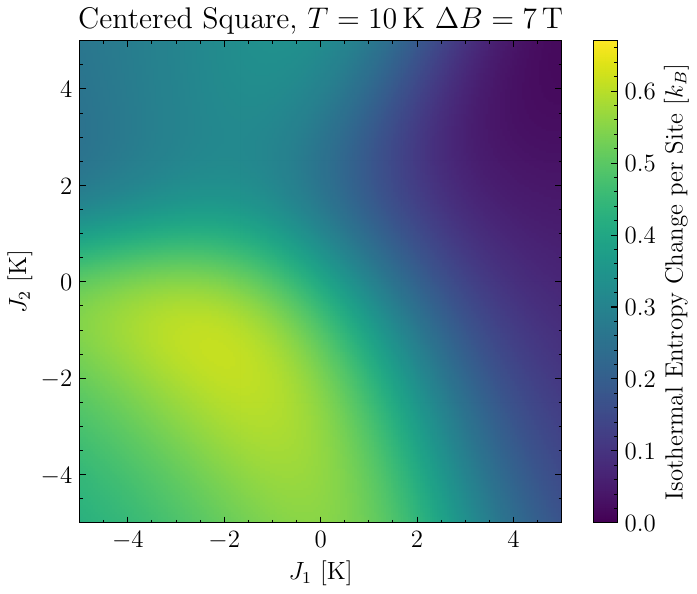}
    \includegraphics[width=0.80\columnwidth]{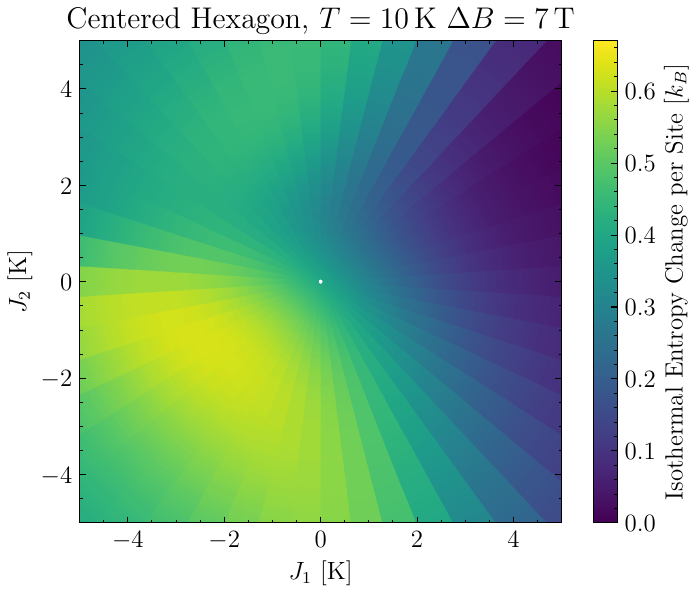}
    \includegraphics[width=0.80\columnwidth]{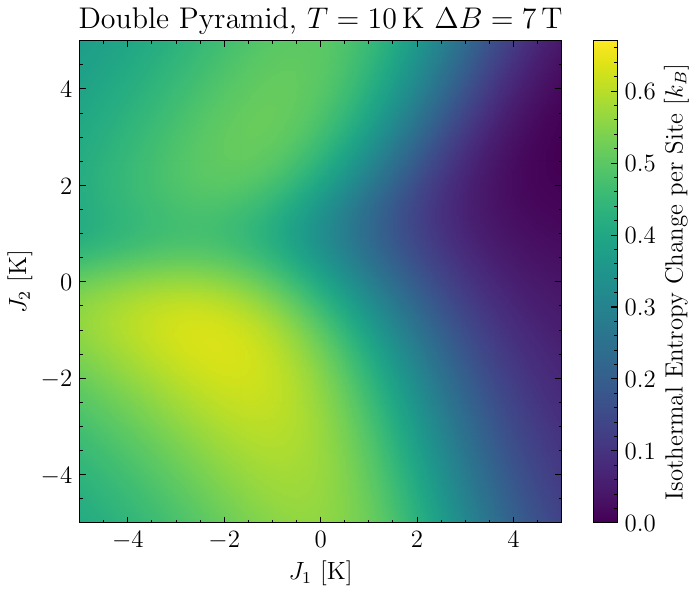}
    \caption{Isothermal entropy change per site when going from $B=\SI{7}{\tesla}$ to $B=0$ at a temperature of \SI{10}{\kelvin} for all three topologies. Each site has spin 3/2.}
    \label{fig:entropy-change-3-3-10K}
\end{figure}

Figure \ref{fig:entropy-change-3-3-10K} shows how $\Delta S(T=\SI{10}{\kelvin},\Delta B=\SI{7}{\tesla})$ changes 
when varying the interaction strengths for all three molecular structures of interest. 
The first striking realization is the qualitative similarity of the results for all studied topologies. 
The main difference is quantitative, with the centered square having a smaller entropy change 
per site than the centered hexagon, and the double pyramid having the biggest of them all. 
Also note that the peak is reached when both interactions are ferromagnetic, 
and the peak is quite flat. This flatness implies that it will be easier for chemists to synthesize 
a molecule that approximately maximises $\Delta S(T=\SI{10}{\kelvin},\Delta B=\SI{7}{\tesla})$.

\begin{figure}
    \centering
    \includegraphics[width=0.80\columnwidth]{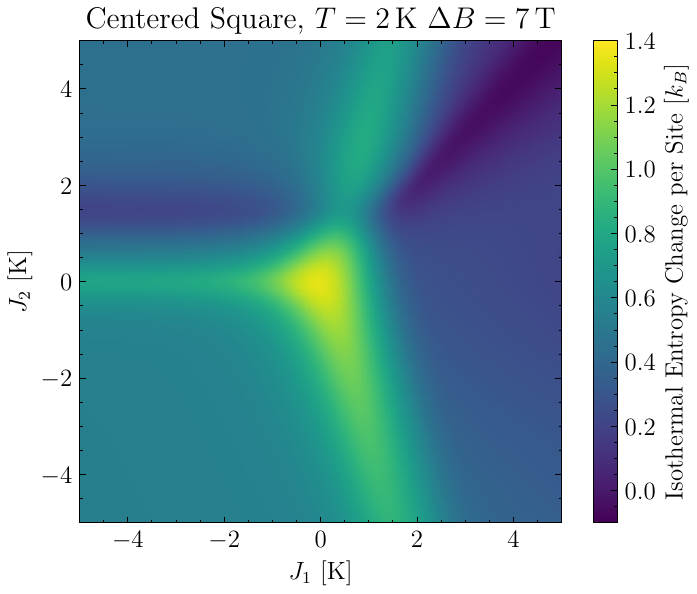}
    \includegraphics[width=0.80\columnwidth]{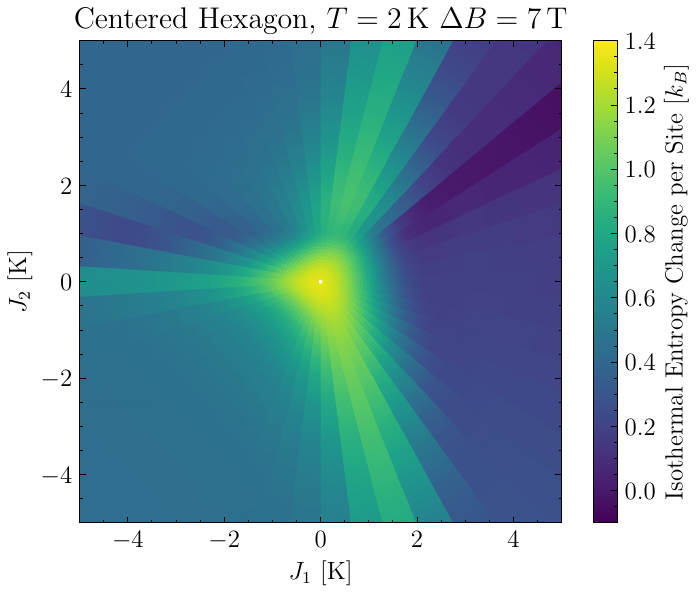}
    \includegraphics[width=0.80\columnwidth]{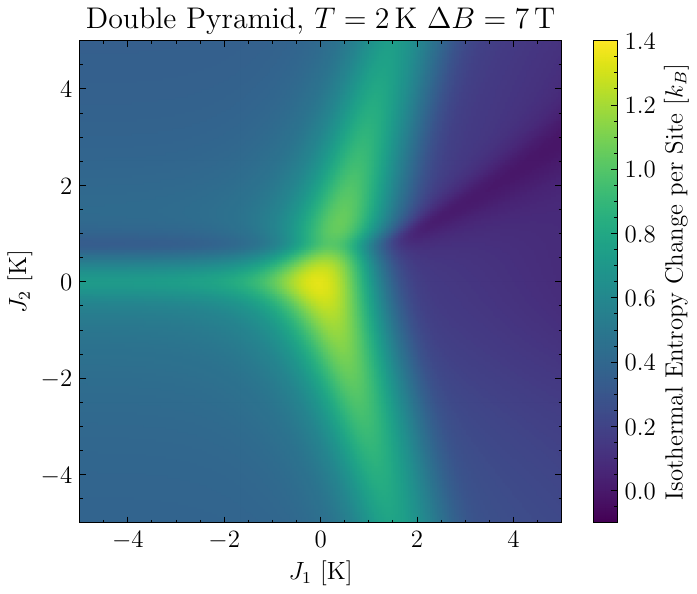}
    \caption{Isothermal entropy change per site when going from $B=\SI{7}{\tesla}$ to $B=0$ at a temperature of \SI{2}{\kelvin} for all three topologies. Each site has spin 3/2.}
    \label{fig:entropy-change-3-3-2K}
\end{figure}

$\Delta S(T=\SI{2}{\kelvin},\Delta B=\SI{7}{\tesla})$ is also qualitatively similar across topologies with the notable exception 
of a band of negative entropy change in the purely antiferromagnetic quadrant, whose slope changes across topologies, as can be seen in \figref{fig:entropy-change-3-3-2K}. The maximum entropy change is reached around the paramagnetic limit, i.e. $J_1 = J_2 = 0$. 
One would expect the entropy change per site in this limit to be the same for all three molecular structures due to the lack of interactions between the sites and the extensivity of entropy. This is verified numerically, 
$\Delta S(J_1 = J_2 = 0;\, T = \SI{2}{\kelvin}) / (\text{\# sites}) \approx 1.4~k_B$ for the three structures and $\Delta B=\SI{7}{\tesla}$.

\begin{figure}
    \centering
    \includegraphics[width=1.0\columnwidth]{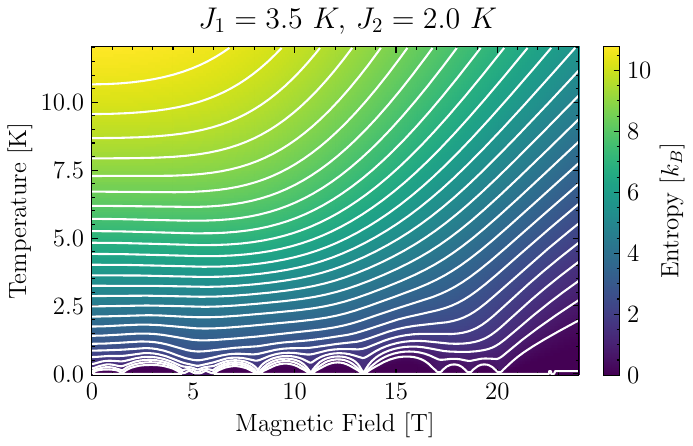}
    \caption{Isentropes of the double pyramid with spins $3/2$ and $J_1 = \SI{3.5}{\kelvin}$ and $J_2 = \SI{2}{\kelvin}$. The point $(J_1,\, J_2)$ lies on the negative band of \figref{fig:entropy-change-3-3-2K}.}
    \label{fig:isentropes-dark}
\end{figure}

The negative band can be explained by inspecting the isentropes. For instance, take the double pyramid with $J_1 = \SI{3.5}{\kelvin}$ and $J_2 = \SI{2}{\kelvin}$, whose isentropes are shown on \figref{fig:isentropes-dark}. Level crossings induce ``hilly" isentropes for low entropy (and hence low temperature). Around \SI{2}{\kelvin}, this results in the isentropes being ``pulled down" and hence the negative entropy change. Notice also that the absolute value of the change is small because in the range $\SI{2}{\kelvin} \lesssim T \lesssim \SI{5}{\kelvin}$ and $B \lesssim \SI{10}{\tesla}$ the isentropes are remarkably flat.

\begin{figure}
    \centering
    \includegraphics[width=1.0\columnwidth]{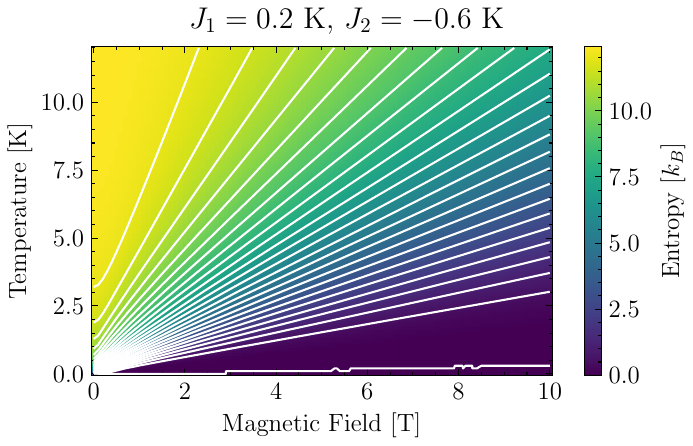}
    \includegraphics[width=1.0\columnwidth]{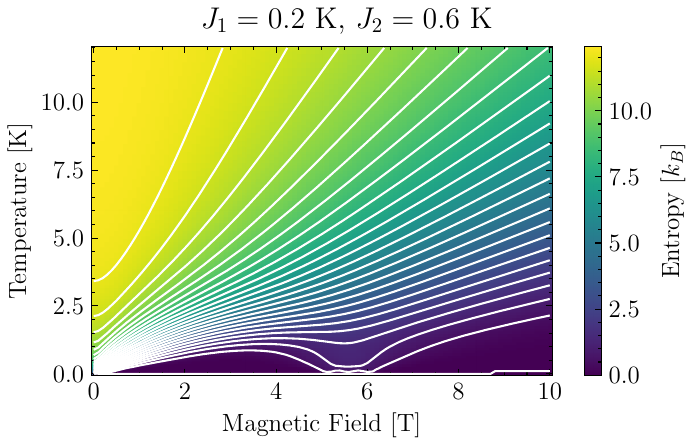}
    \includegraphics[width=1.0\columnwidth]{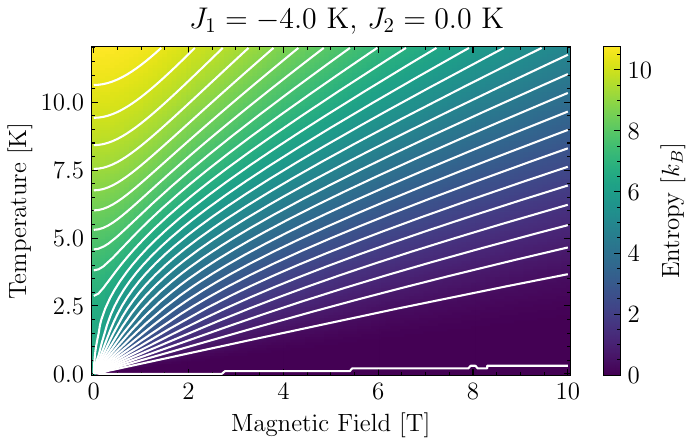}
    \caption{Isentropes of the double pyramid with spins $3/2$
     for interaction strengths $(J_1,\, J_2)$ lying on the three yellow branches in \figref{fig:entropy-change-3-3-2K}. More specifically, the sub-figures from top to bottom correspond to: the upper branch (quadrant I); the lower branch (quadrant IV); and the left branch (negative $J_1$-axis).
    }
    \label{fig:isentropes-yellow}
\end{figure}

The yellow bands of \figref{fig:entropy-change-3-3-2K} can also be explained by looking at isentropes. \figref{fig:isentropes-yellow} shows the isentropes for the three yellow bands for the example of the double-pyramid. The isentropes crossing the neighborhood of $B = \SI{7}{\tesla},\, T = \SI{2}{\kelvin}$ are close to those of a paramagnet, i.e. straight lines passing through the origin, which explains why they behave similarly to the paramagnetic case $J_1 = J_2 = 0$.

\begin{figure}
    \centering
    \includegraphics[width=0.80\columnwidth]{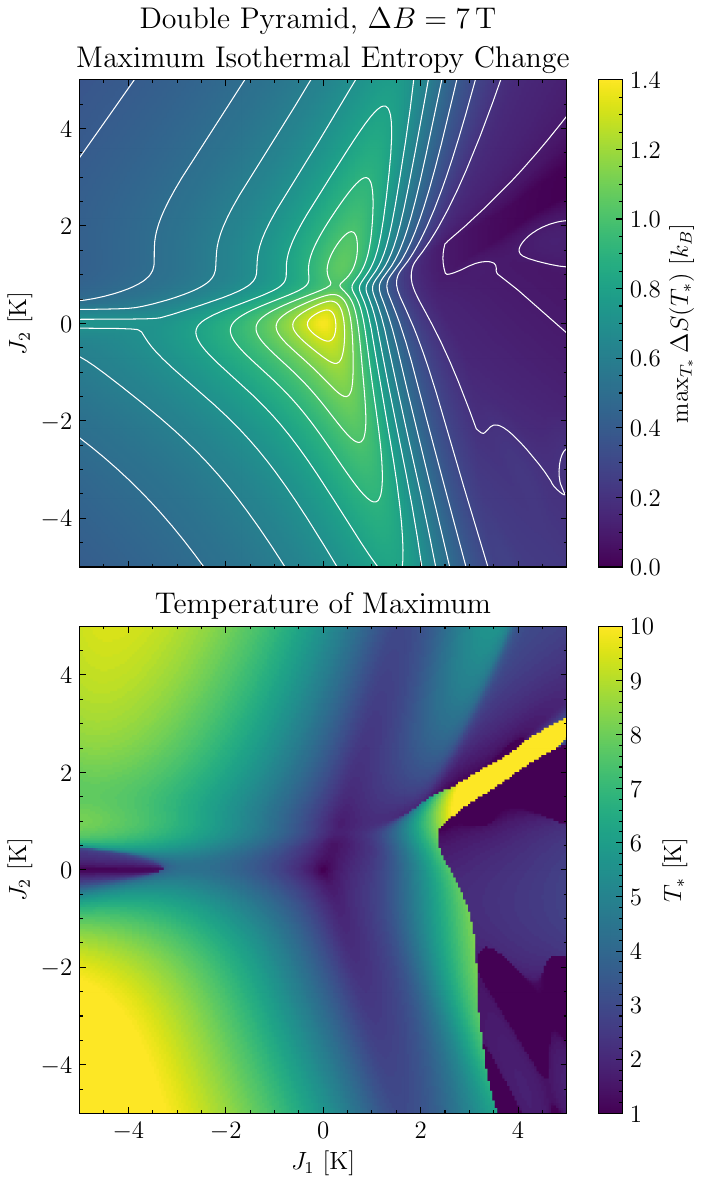}
    \caption{Maximum isothermal entropy change per site $\max_{T_*} \Delta S(T_*)$ for different values of $(J_1, J_2)$ and the temperature $T_*$ at which it is realized. Note that the search has been restrained to the interval $T_* \in [\SI{1}{\kelvin}, \SI{10}{\kelvin}]$. The contour lines range from 0.1~$k_B$ to 1.3~$k_B$ and differ by increments of 0.1~$k_B$. The system studied is the double-pyramid with spins $3/2$.}
    \label{fig:pyramid-3-3-max-entropy}
\end{figure}

Since lowering the temperature increases $\Delta S$ for some $(J_1, J_2)$ and decreases it for some others, 
it is natural to wonder which temperature $T_*$ maximizes the isothermal entropy change for some given interaction strengths. 
This information and the value of the maximum is illustrated in \figref{fig:pyramid-3-3-max-entropy} for the double pyramid. 
Note that since the search is restricted to the interval $T_* \in [\SI{1}{\kelvin}, \SI{10}{\kelvin}]$, 
it is possible that in regions where $T_* = \SI{1}{\kelvin}$ ($T_* = \SI{10}{\kelvin}$) the actual 
value of $T_*$ is $< \SI{1}{\kelvin}$ ($> \SI{10}{\kelvin}$). The maximum $\Delta S$ is reached in
the paramagnetic limit and possibly when $T_* < \SI{1}{\kelvin}$. 
The figure also shows that ferromagnetic interactions maximize the entropy change when the 
temperature is $\gtrsim \SI{5}{\kelvin}$.

\begin{figure}
    \centering
    \includegraphics[width=\columnwidth]{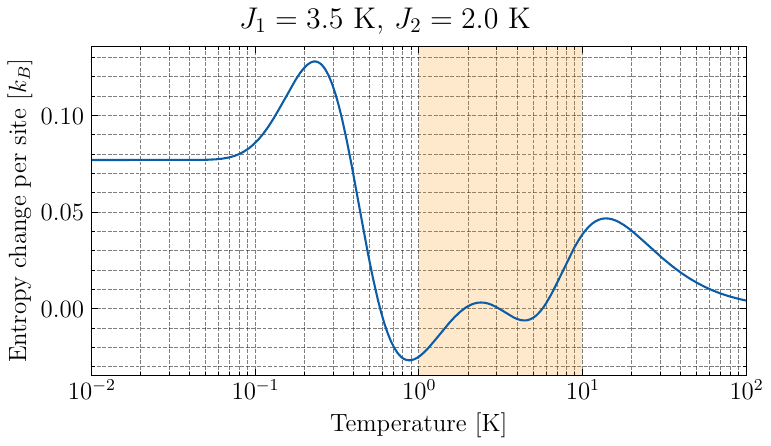}
    \caption{Isothermal entropy change per site with respect to temperature. The system studied is the double pyramid with spins $3/2$, $J_1 = \SI{3.5}{\kelvin}$, $J_2 = \SI{2.0}{\kelvin}$, where the magnetic field was changed from \SI{7}{\tesla} to 0. The original search interval for the maximum is highlighted.}
    \label{fig:ds-vs-temp-3.5-2}
\end{figure}

The discontinuity of $T_*$ in \figref{fig:pyramid-3-3-max-entropy} is due to the restriction $T_* \in [\SI{1}{\kelvin}, \SI{10}{\kelvin}]$. For instance, for $J_1 = \SI{3.5}{\kelvin}$ and $J_2 = \SI{2}{\kelvin}$, it would seem the maximum isothermal entropy change is reached at $T_* = \SI{10}{\kelvin}$. However plotting $\Delta S(T)$ for these interaction strengths on a wider temperature range (\figref{fig:ds-vs-temp-3.5-2}) shows that entropy change is maximized at $T = \SI{0.23}{\kelvin}$. Note however that the restriction $T_* \in [\SI{1}{\kelvin}, \SI{10}{\kelvin}]$ still makes sense because we are interested in possible applications to sub-Kelvin cooling starting from a temperature in this range, not in the characterization of the materials.

\subsubsection{\texorpdfstring{$T_\text{cold}$}{T_cold}}
\label{sec:t-cold}

\begin{figure}
    \centering
    \includegraphics[width=0.80\columnwidth]{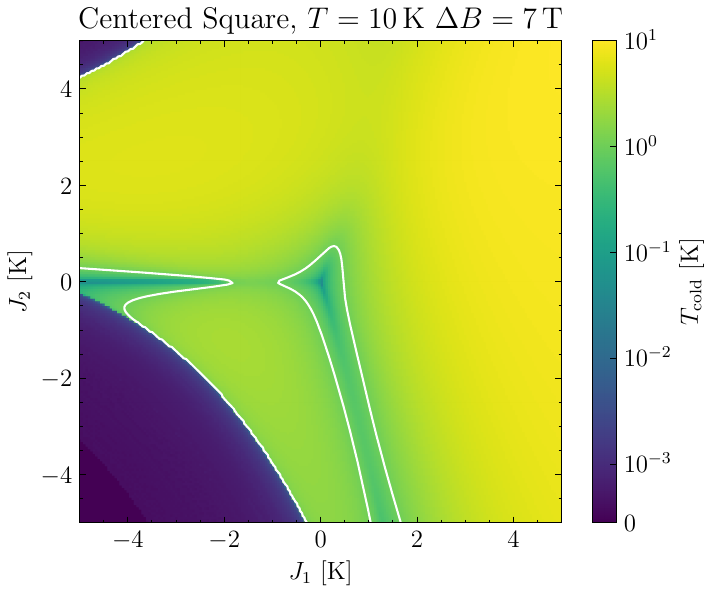}
    \includegraphics[width=0.80\columnwidth]{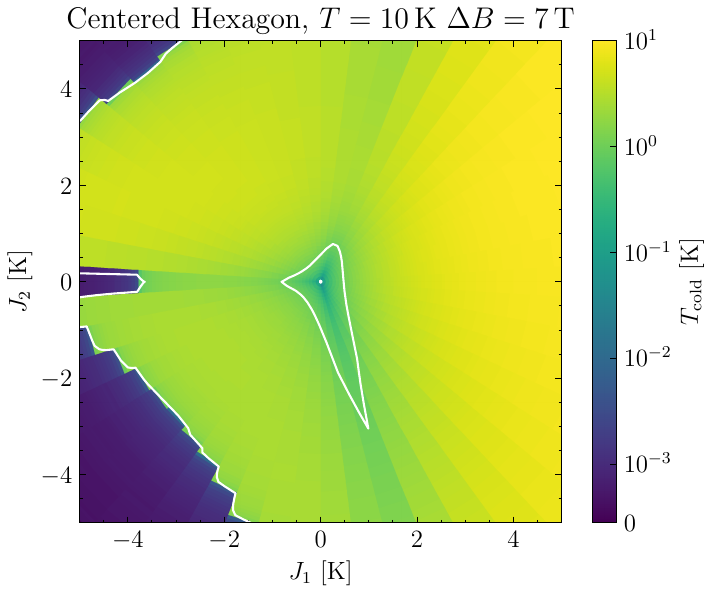}
    \includegraphics[width=0.80\columnwidth]{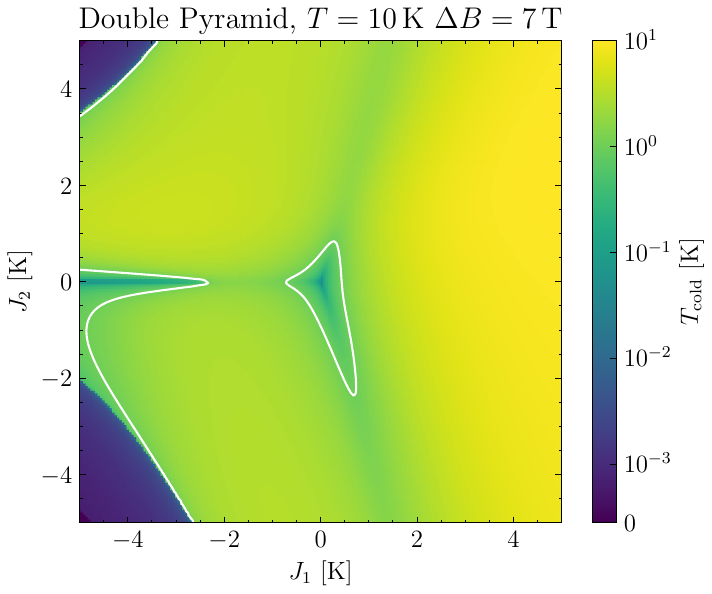}
    \caption{Temperature $T_\text{cold}$ reached after adiabatic demagnetization (from $B = \SI{7}{\tesla}$ to $B=0$) when the samples starts at \SI{10}{\kelvin} for all three topologies. All sites have spin 3/2. The contour line $T_\text{cold} = \SI{1}{\kelvin}$ is shown in white.}
    \label{fig:t-cold-3-3-10K}
\end{figure}

The temperature $T_\text{cold}$ reached after adiabatic demagnetization of a sample initially at \SI{10}{\kelvin} is shown on \figref{fig:t-cold-3-3-10K}. Once again, all topologies have qualitatively the same behaviour, with the notable exception of the central sub-Kelvin region of the centered square. It has a greater extent, and it is also unclear whether the region is bound\HM{ed} or whether it extends infinitely in the fourth quadrant. There is a sudden drop in $T_\text{cold}$ in the third quadrant (from \SI{1}{\kelvin} to \SI{3}{\milli\kelvin} for the double pyramid). The peak of entropy change is inside of this zone of sub-Kelvin cooling for the centred square, but not for the centered hexagon or the double pyramid. The other regions that reach below \SI{1}{\kelvin} correspond to fine-tuned Hamiltonians that would be impractical to realize in real molecules. 

\begin{figure}
    \centering
    \includegraphics[width=0.80\columnwidth]{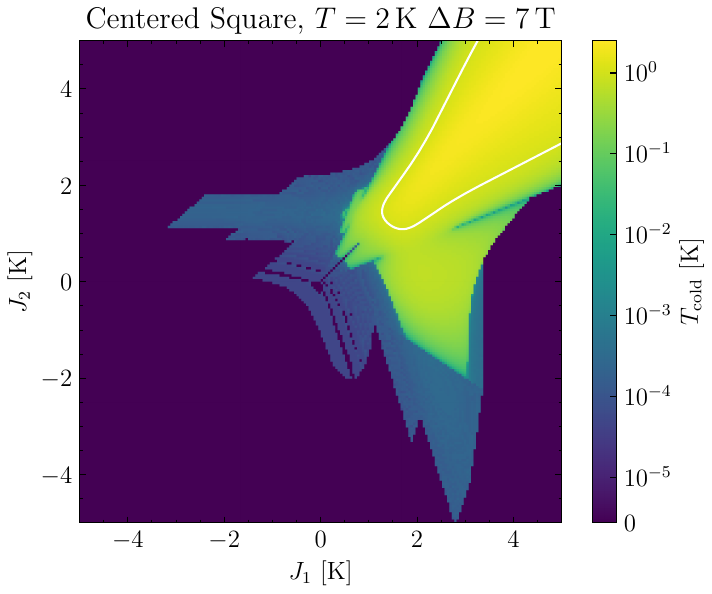}
    \includegraphics[width=0.80\columnwidth]{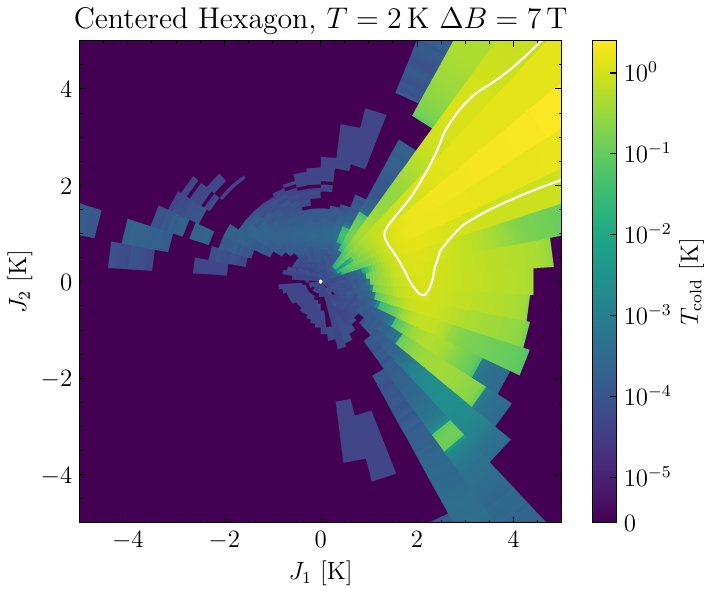}
    \includegraphics[width=0.80\columnwidth]{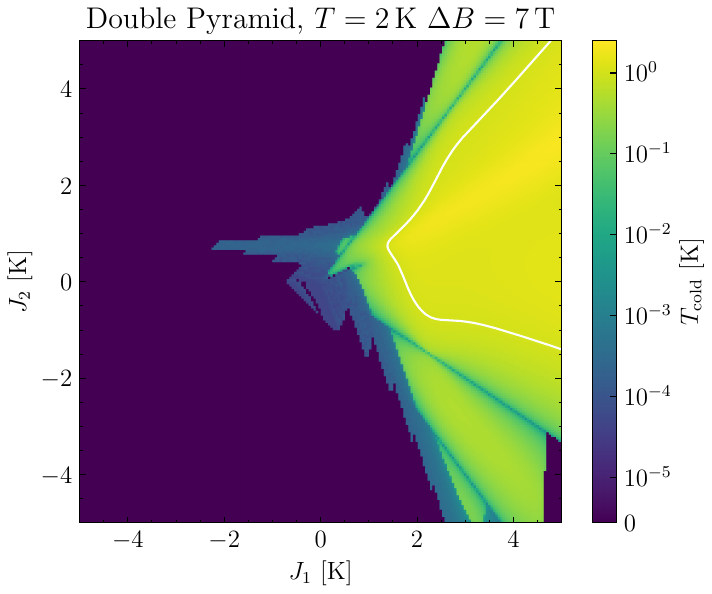}
    \caption{Temperature $T_\text{cold}$ reached after adiabatic demagnetization (from $B = \SI{7}{\tesla}$ to $B=0$) when the samples starts at \SI{2}{\kelvin} for all three topologies. All sites have spin 3/2. The contour line $T_\text{cold} = \SI{1}{\kelvin}$ is shown in white.}
    \label{fig:t-cold-3-3-2K}
\end{figure}

Figure \ref{fig:t-cold-3-3-2K} shows $T_\text{cold}(J_1, J_2; T = \SI{2}{\kelvin})$ for the three molecular structures. 
Sub-kelvin cooling is achieved if $J_1$ is ferromagnetic, no matter $J_2$. The paramagnetic limit also achieves such cooling. 
The jaggedness of the borders between the regions where $T_\text{cold} = \SI{0}{\kelvin}$ and $T_\text{cold} > \SI{0}{\kelvin}$ 
is probably due to numerical inaccuracies of the numerical search program 
and due to the fact that $S(T, B=0)$ is flat near $T=\SI{0}{\kelvin}$.

\subsection{Effects of a lower central spin}
\label{sec:lower-central-spin}

\begin{figure}
    \centering
    \includegraphics*[width=0.80\columnwidth]{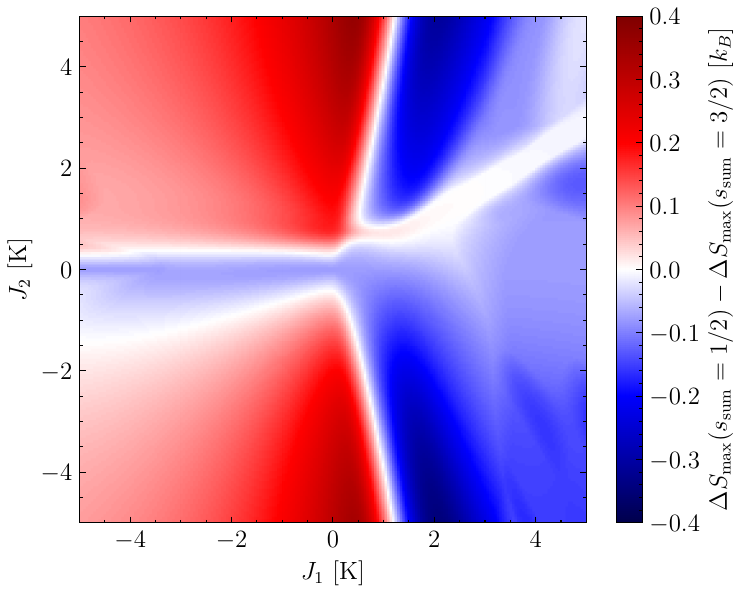}
    \caption{Difference in $\Delta{}S_\text{max}$ between the cases of $s_{\text{summit}} = 3/2$ and $s_{\text{summit}} = 1/2$ for the double pyramid. Red (blue) indicates that reducing $s_\text{summit}$ increases (decreases) $\Delta{}S_\text{max}$. The entropy is given per number of sites.}
    \label{fig:pyramid-diff-max-entropy}
\end{figure}

In this section, we study how $\Delta S_\text{max} / N$, where $N = 9$ is the number of sites, 
changes in the double pyramid when lowering the central spin, which can also be thought of as the shared 
``summit" of the two pyramids. More precisely, we compare the case where all sites have spin 3/2 and 
the case where the summit has $s_\text{summit} = 1/2$ and all the other sites have spin 3/2. 
The general rule of thumb is that larger spins are preferential for a bigger magneto-caloric effect, 
however it is clear from \figref{fig:pyramid-diff-max-entropy} that this is not necessarily the case. 
Reducing $s_\text{summit}$, and hence the maximal total spin of the molecule, actually increases 
$\Delta S_\text{max}$ if $J_1$ is ferromagnetic unless $\SI{-1.5}{\kelvin} \lesssim J_2 \lesssim 0$, 
at least for the range of $(J_1, J_2)$ that has been explored. 
Also note that this increase is not negligible. Take for instance 
$J_1 = J_2 = \SI{-2}{\kelvin}$; $\Delta S_\text{max}(s_\text{summit} = 3/2) / N = 0.69k_B$, 
and $\Delta S_\text{max}(s_\text{summit} = 1/2) / N = 0.76k_B$, which is an increase 
of $0.07k_B$ or approximately \SI{10}{\percent}.

This effect can be understood by looking at the classical spin structure.
For ferromagnetic $J_1$ and antiferromagnetic $J_2$ the spins of the squares are antiparallel 
to the central spin. If the latter is smaller in magnitude the total (ground state) spin 
of this configuration is bigger, compare \cite{MHS:DT14,GHG:CCR15}. 
This leads to more low-lying levels which enhances the magnetocaloric effect at low temperatures.

\section{Summary}
\label{sec:discussion}

In this paper, we investigate the molecular magnetic properties of a class of centered 
magnetic molecules (the centered square, the centered hexagon as well as the double 
pyramid). All three of them can be modelled using two exchange constants, and they
exhibit very similar properties. Their low-lying densities of state can be largely
modified by varying $J_1$ and $J_2$, and this reflects in the broad spectrum of magnetocaloric 
properties.

From the molecular perspective, ferromagnetic interactions seem to be preferential 
for application. However, if the material orders at temperatures above 
the desired $T_\text{cold}$, this would preclude further cooling. Then, a compromise might be beneficial
and frustrated antiferromagnetic configurations with a large low-lying density of states
might turn out more useful. In addition, the common believe that larger spins
are always preferential does not hold. In particular, in ferrimagnetic arrangements
it can be advantageous that some spins have a small spin quantum number.

\section*{Acknowledgment}

This work has received support from the EU via MSCA-DN MolCal, 101119865. 
We thank Henrik Dick for explaining the method used to find the 
analytical solutions in a group seminar.

\appendix

\section{Analytical Solutions}
\label{sec:appendix-analytical-solutions}

Both the centered square and double pyramid can be analytically solved by "completing the square". The basic idea of this method is to rewrite exchange interactions as $\qvecspin{i} \cdot \qvecspin{j} = \left(\qvecspin{ij}^2 - \qvecspin{i}^2 - \qvecspin{j}^2\right)/2$, where for brevity's sake we wrote $\qvecspin{ij} := \qvecspin{i} + \qvecspin{j}$. The fact that the result only depends on squared spin operators nudges one into finding a coupling scheme such that the different operators are independent of each other. A method to determine if such as solution is possible, and find it if so, is presented in \cite{StS:MPAG09} and elaborated in a group seminar by Henrik Dick.

\subsection{Centered Square}

\begin{figure}[H]
    \centering
    \includegraphics*[width=0.80\columnwidth]{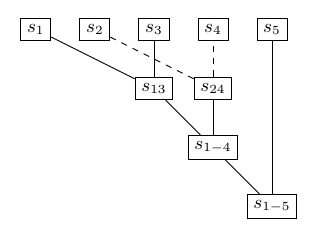}
    \caption{Coupling scheme for the centered square.}
    \label{coupling-scheme-centered-square}
\end{figure}

Label the sites of the square frame from 1 to 4 and the central as the 5\textsuperscript{th} site. 
The total spin quantum number at site $i$ is written $s_i$ and $s_{ijk\dots}$ is the 
total spin quantum number for the coupling of the spins at sites $i$, $j$, $k$, and so on. 
The coupling scheme is described in \figref{coupling-scheme-centered-square}. 
The quantum numbers indexing the energy eigenstates are 
$s_{13}$, $s_{24}$, $s_{1-4}$, $s_{1-5}$, and $M$, 
which is the eigenvalue of $\qop{S}_{1-5}^z$, where $\qvecspin{i-j} := \sum_{k=i}^j \qvecspin{k}$.
The Hamiltonian can be written as
\begin{multline}
        \qop{H} = \frac{J_2}{2}\left(\qvecspin{1-5}^2 - \qvecspin{5}^2\right) + \frac{J_1 - J_2}{2}\qvecspin{1-4}^2 \\
        - \frac{J_1}{2}\left(\qvecspin{13}^2 + \qvecspin{24}^2\right) + g \mu_B B \qop{S}_{1-5}^z
        \ .
\end{multline}

\vspace*{1mm}

\subsection{Double Pyramid}

\begin{figure}[H]
    \centering
        \includegraphics*[width=0.80\columnwidth]{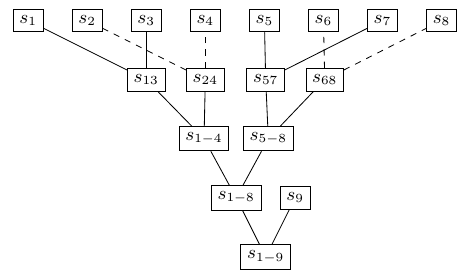}
    \caption{Coupling scheme for the double pyramid.}
    \label{coupling-scheme-double-pyramid}
\end{figure}

Label the sites of one of the square basis from 1 to 4, the sites of the other basis from 5 to 8, and the summit as the 9\textsuperscript{th} site. Couple the spins according to \figref{coupling-scheme-double-pyramid}. 
The Hamiltonian can then be expressed as
\begin{multline}
    \qop{H} = \frac{J_1}{2} \left(\qvecspin{1-4}^2 - \qvecspin{13}^2 - \qvecspin{24}^2 + \qvecspin{5-8}^2 - \qvecspin{57}^2 - \qvecspin{68}^2\right) \\
    + \frac{J_2}{2} \left(\qvecspin{1-9}^2 - \qvecspin{1-8}^2 - \qvecspin{9}^2\right) + g \mu_B B \qop{S}_{1-9}^z
    \ .
\end{multline}

%

\end{document}